\documentclass[a4paper,12pt]{article}
\usepackage{amsmath}
\usepackage{amssymb}
\usepackage{amsfonts}
\usepackage{bbm} 
\usepackage{fleqn} 
\usepackage[footnotesize]{caption}
\usepackage{braket} 
\usepackage{mathrsfs}

\def\da{{\delta}_1}
\def\db{{\delta}_2}
\def\dc{{\delta}_3}

\def\daP{{\delta}'_1}
\def\dbP{{\delta}'_2}
\def\dcP{{\delta}'_3}

\def\<{\left\langle}
\def\>{\right\rangle}
\def\ChargeC{\mathrm{C}}

\allowdisplaybreaks[2]

\begin{document}

\bibliographystyle{OurBibTeX}

\begin{titlepage}

 \vspace*{-15mm}
\begin{flushright}
SHEP/406
\end{flushright}
\vspace*{5mm}

\begin{center}
{
\sffamily\LARGE
Neutrino Mixing from the Charged Lepton Sector \\[1mm]
with Sequential Right-Handed Lepton Dominance 
}
\\[12mm]
S. Antusch\footnote{E-mail: \texttt{santusch@hep.phys.soton.ac.uk}},
S. F. King\footnote{E-mail: \texttt{sfk@hep.phys.soton.ac.uk}}
\\[1mm]
{\small\it
Department of Physics and Astronomy,
University of Southampton,\\
Southampton, SO17 1BJ, U.K.
}
\end{center}
\vspace*{1.00cm}

\begin{abstract}

\noindent We systematically analyze the possibility that bi-large lepton 
mixing originates from the charged lepton sector 
in models with sequential dominance for the right-handed 
charged leptons.  
We derive analytical expressions for the mixing 
angles and CP phases of the MNS matrix for the case of zero mixing
from the neutrino sector, which is arranged for with the help of 
sequential dominance for the right-handed neutrinos.  
For small $\theta_{13}$, the two large mixing angles $\theta_{12}$ and 
$\theta_{23}$ are determined by the Yukawa couplings to the 
dominant right-handed tau. The mixing angle $\theta_{13}$ is then governed
 by the subdominant right-handed muon Yukawa couplings. 
Naturally small $\theta_{13}$ and sequential right-handed lepton dominance 
can be realized in type I see-saw models and their type II upgrades via  
spontaneously broken SO(3) flavour symmetry and real vacuum alignment.        
We discuss the prediction for $\theta_{13}$ in this scenario 
and corrections to it including the dependence on the neutrino mass scale. 
\end{abstract}

\end{titlepage}
\newpage
\setcounter{footnote}{0}

\section{Introduction}
The mixing matrix in the lepton sector, the MNS matrix $U_{\mathrm{MNS}}$, 
is defined by the charged electroweak current 
$\overline{{e}_\mathrm{L}}^f \gamma^\mu  U_{MNS} {\nu}^f_\mathrm{L}$ 
in the mass basis. 
It is of course possible to choose a basis where all the 
lepton mixing stems entirely from the neutrino sector or 
from the charged lepton sector. 
From the perspective of model building, the goal is to identify
symmetries and mechanisms for understanding the observed fermion masses and 
mixings. Here it makes a big difference whether the mixing is generated in 
the neutrino sector, the charged lepton sector or maybe partly in both 
sectors. For a review on neutrino mass models, see e.g.~\cite{King:2003jb}.      
Experimentally, it has been found that the lepton 
mixing angles $\theta_{12}$ and $\theta_{23}$ are both large whereas 
$\theta_{13}$ is comparably small. 

Recently, there has been some interest in the possibility of generating 
the large neutrino mixings in the charged lepton sector. 
In \cite{Altarelli:2004jb,Romanino:2004ww}, bi-large mixing and small 
$\theta_{13}$ from the charged lepton mass matrix have been achieved 
by a kind of see-saw mechanism in the charged lepton sector. 
In an earlier study \cite{Babu:2001cv}, a texture for 
bi-large mixing and small $\theta_{13}$ from the charged leptons 
and its realization by effective operators in the SO(10)-GUT framework 
have been considered. 

In this work, we systematically analyze the possibility that the 
lepton mixing originates
 from the charged lepton sector in models with sequential 
dominance for the {\em right-handed charged leptons} 
as well as for right-handed 
neutrinos \cite{King:1999mb,King:2002nf}. 
We find that for several reasons, 
such a generalized sequential dominance is a useful foundation for models of this type: 
It implies the desired hierarchy for the charged lepton masses 
and small right-handed charged lepton mixings. 
In addition, it can naturally lead to small mixings from the neutrino sector, 
even if the neutrino Yukawa matrix has a similar structure to the charged 
lepton Yukawa matrix.  
For investigating the requirements for bi-large mixing with naturally small 
$\theta_{13}$, we derive analytic expressions for 
the mixing angles and CP phases of the MNS matrix for the case of 
zero mixing coming from the neutrino sector. 
We find that for small $\theta_{13}$, the two large mixing angles $\theta_{12}$ and 
$\theta_{23}$ are governed by the Yukawa couplings to the 
dominant right-handed tau. The mixing angle $\theta_{13}$ is then determined by 
the subdominant right-handed muon Yukawa couplings. 
We find a condition for naturally small $\theta_{13}$, which is different 
from the vanishing determinant condition of 
\cite{Altarelli:2004jb,Romanino:2004ww}.

Sequential right-handed lepton dominance and lepton mixing originating 
 from the charged lepton sector
can be realized via  
spontaneously broken SO(3) flavour symmetry and real vacuum alignment 
\cite{Antusch:2004xd}.  
Three zero entries in the neutrino and charged lepton Yukawa couplings, 
leading to small mixing from the neutrino sector in combination 
with small lepton mixing $\theta_{13}$,  
can arise with real alignment of the SO(3)-breaking vacuum.  
Type I see-saw models of this type can be upgraded to 
type II see-saw  models,  
where hierarchical 
neutrino masses are as natural a partially degenerate ones. 
We discuss the prediction for $\theta_{13}$ in this scenario 
and how corrections to it depend on the neutrino mass scale.

\section{Our Conventions}
For the mass matrix of the charge leptons $M_e=Y_e v_d$ defined by 
$\mathcal{L}_e=-M_e \overline e^f_{\mathrm{L}} e^f_{\mathrm{R}}$ + h.c. 
and for a neutrino mass
matrix $m_{LL}^\nu$ of Majorana-type defined by  
$\mathcal{L}_\nu=-\tfrac{1}{2} m_{LL}^\nu \overline \nu_{\mathrm{L}} 
\nu^{\ChargeC f}_{\mathrm{L}}$ + h.c., 
the change from flavour basis to mass
eigenbasis can be performed with the unitary diagonalization matrices 
$U_{e_\mathrm{L}},U_{e_\mathrm{R}}$ and 
$U_{\nu_\mathrm{L}}$ by  
\begin{eqnarray}\label{eq:DiagMe}
U_{e_\mathrm{L}} \, M_e \,U^\dagger_{e_\mathrm{R}} =
\left(\begin{array}{ccc}
\!m_e&0&0\!\\
\!0&m_\mu&0\!\\
\!0&0&m_\tau\!
\end{array}
\right)\! , \quad
U_{\nu_\mathrm{L}} \,m^\nu_{\mathrm{LL}}\,U^T_{\nu_\mathrm{L}} =
\left(\begin{array}{ccc}
\!m_1&0&0\!\\
\!0&m_2&0\!\\
\!0&0&m_3\!
\end{array}
\right)\! .
\end{eqnarray}  
The MNS matrix is then given by
\begin{eqnarray}
U_{\mathrm{MNS}} = U_{e_\mathrm{L}} U^\dagger_{\nu_\mathrm{L}}\; .
\end{eqnarray}
We use the parameterization 
$
U_{\mathrm{MNS}} = R_{23} U_{13} R_{12} P_0
$ 
with $R_{23}, U_{13}, R_{12}$ and $P_0$ being defined as 
\begin{align}\label{eq:R23U13R12P0}
R_{12}:=
\left(\begin{array}{ccc}
  c_{12} & s_{12} & 0\\
  -s_{12}&c_{12} & 0\\
  0&0&1\end{array}\right)
  , \:&
\quad U_{13}:=\left(\begin{array}{ccc}
   c_{13} & 0 & s_{13}e^{-i\delta}\\
  0&1& 0\\
  - s_{13}e^{i\delta}&0&c_{13}\end{array}\right)  ,  \nonumber \\ 
R_{23}:=\left(\begin{array}{ccc}
 1 & 0 & 0\\
0&c_{23} & s_{23}\\
0&-s_{23}&c_{23}
 \end{array}\right)
  , \:&
 \quad P_0:= 
 \begin{pmatrix}
 1&0&0\\0&e^{i\beta_2}&0\\0&0&e^{i\beta_3}
  \end{pmatrix}  
\end{align}
and where $s_{ij}$ and $c_{ij}$ stand for $\sin (\theta_{ij})$ and $\cos
(\theta_{ij})$, respectively. 
The matrix $P_0$ contains the possible Majorana 
phases $\beta_2$ and $\beta_3$.  
 $\delta$ is the Dirac CP phase relevant for neutrino oscillations.

\section{Sequential Dominance for Right-Handed Leptons}\label{sec:SmallNuMixing}
We first generalize the notion of sequential right-handed neutrino dominance 
(RHND) \cite{King:1999mb,King:2002nf}, which is an extension of single 
right-handed neutrino dominance \cite{King:1998jw,King:1999cm}, to all
right-handed leptons. 
Let us therefore write the Yukawa
couplings for the charged leptons and the neutrinos $Y_e$ and
$Y_\nu$ and the heavy Majorana mass matrix for the charged leptons as 
\begin{eqnarray}
Y_e = \left(
\begin{array}{ccc}
p & d & a \\
q & e & b \\
r & f & c
\end{array}
\right)\!,\;\;
Y_\nu = \left(
\begin{array}{ccc}
p' & d' & a' \\
q' & e' & b' \\
r' & f' & c'
\end{array}
\right)\!,\;\;
M_\mathrm{RR}
= 
\left(\begin{array}{ccc}
\!M_\mathrm{R1}&0&0\!\\
\!0&M_\mathrm{R2}&0\!\\
\!0&0&M_\mathrm{R3}\!
\end{array}
\right)\!.
\end{eqnarray}
We can choose a basis where 
$M_\mathrm{R1},M_\mathrm{R2}$ and $M_\mathrm{R3}$ are real and positive. 
The entries in the Yukawa matrices are in
general complex.   
In the case of a type I see-saw 
scenario 
\cite{Yanagida:1980,Glashow:1979vf,Gell-Mann:1980vs,Mohapatra:1980ia}, 
the neutrino mass matrix is given by
\begin{eqnarray}
m^\nu_{\mathrm{LL}} = m^\mathrm{I}_{\mathrm{LL}} = - v_u^2\, Y_\nu M^{-1}_\mathrm{RR} Y^T_\nu \; .
\end{eqnarray} 
 We will assume sequential dominance among the right-handed neutrinos 
 and charged leptons, respectively. In our notation, 
each right-handed charged lepton couples to a column in $Y_e$ and 
each right-handed neutrino to a column in $Y_\nu$. 
For the charged leptons, the sequential dominance conditions are
\begin{eqnarray}\label{eq:SeqDominanceCL}
|a|,|b|,|c| \gg |d|,|e|,|f| \gg |p|,|q|,|r| \;.
\end{eqnarray}
They imply the desired hierarchy for the charged lepton masses $m_\tau \gg m_\mu
\gg m_e$ and small right-handed mixing of $U_{e_\mathrm{R}}$. In the neutrino sector, the sequential 
dominance conditions determine which
column of $Y_\nu$ in combination with the corresponding 
mass eigenvalues of the right-handed
neutrinos $M_{\mathrm{R}i}$ 
contributes dominantly and which sub-dominantly to $m^\nu_{\mathrm{LL}}$. 
Usually, sequential RHND 
\cite{King:1999mb,King:2002nf} is viewed as a framework 
for generating large solar 
mixing $\theta_{12}$ and large atmospheric mixing $\theta_{23}$ in the neutrino 
 mass matrix. 
However, given sequential dominance in the neutrino sector, 
one can easily find the conditions for small mixing from the neutrinos as well. 
Small mixing from the neutrino sector requires three small entries in $Y_\nu$. 
In particular, 
we shall require 
$Y_\nu$ and $M_{\mathrm{RR}}$ to have the approximate form 
\begin{eqnarray}\label{eq:SmallMixingFromMnu}
Y_\nu \approx \left(
\begin{array}{ccc}
a' & 0 & 0 \\
b' & q' & 0 \\
c' & r' & f'
\end{array}
\right),\;\;
M_\mathrm{RR}
\approx  
\left(\begin{array}{ccc}
\!X'&0&0\!\\
\!0&X&0\!\\
\!0&0&Y\!
\end{array}
\right),
\end{eqnarray}
with the sequential dominance conditions being 
\begin{eqnarray}\label{eq:SmallMixingFromMnu_seqdom}
\frac{|f'|^2}{Y}\gg 
\frac{|q'|^2,|r'|^2}{X} \gg
\frac{|a'|^2,|b'|^2,|c'|^2}{X'}
\; .
\end{eqnarray}
Small entries in $Y_\nu$ have been neglected. 
Note that equations (\ref{eq:SmallMixingFromMnu}) and (\ref{eq:SmallMixingFromMnu_seqdom}) 
are completely general, since 
a simultaneous reordering of the columns of $Y_\nu$ and of $Y,X$ and $X'$  
leads to the same type I neutrino mass matrix.  
As shown in \cite{Antusch:2004xd},  
three zero entries in $Y_\nu$ might stem from a 
spontaneously broken SO(3) flavour symmetry and real vacuum alignment. 
Other realizations might by found via Abelian or discrete symmetries. 
Note that the the structure of equation (\ref{eq:SmallMixingFromMnu}) 
implies $\theta^\nu_{12}=\theta^\nu_{13}=0$ and a small mixing 
 $\theta^\nu_{23}= {\cal O} (m_2^{\mathrm{I}}/m_3^{\mathrm{I}})$ from the
 neutrino mass matrix. $m_2^{\mathrm{I}}$ and $m_3^{\mathrm{I}}$ are 
 the neutrino mass eigenvalues in a type I
 see-saw model. For a hierarchical mass spectrum, they correspond to the 
 square roots of the solar and atmospheric mass squared differences.  
  We will analyze the consequences of this perturbation for 
  the total lepton mixing angles $\theta_{13},\theta_{12}$ and $\theta_{23}$ 
 in an example in section \ref{sec:C2}.  
  The small $\theta^\nu_{23}$ will in particular contribute to the 
corrections for the $\theta_{13}$-mixing generated in the lepton sector, 
 which we will discuss in section \ref{sec:theta13}.

\section{Bi-Large Mixing from the Charged Lepton Sector}
We now derive analytic expressions for the mixing angles and CP phases of the MNS
matrix in the limit of zero mixing originating from the neutrino sector.
We assume sequential right-handed charged lepton
dominance $|a|,|b|,|c| \gg |d|,|e|,|f| \gg |p|,|q|,|r|$, as in equation (\ref{eq:SeqDominanceCL}).
 Zero mixing from the neutrino sector corresponds to 
\begin{eqnarray}
U_{\nu_{\mathrm{L}}}=  \begin{pmatrix}
 1&0&0\\0&e^{-i\beta^\nu_2}&0\\0&0&e^{-i\beta^\nu_3}
  \end{pmatrix}\!,
\end{eqnarray}
and the MNS matrix is then given by $U_\mathrm{MNS}=U_{e_{\mathrm{L}}}\!\!\cdot
\mbox{diag}\,(1,e^{i \beta^{\nu}_2},e^{i \beta^{\nu}_3})$.    
Note that sequential right-handed charged lepton
dominance implies small right-handed 
mixings of the unitary diagonalization matrix $U_{e_\mathrm{R}}$ defined in
equation (\ref{eq:DiagMe}) and, in particular, 
hierarchical charged lepton masses. 
With sequential right-handed charged lepton dominance,   
the lepton mixing matrix $U_\mathrm{MNS}$ is determined by the requirement 
that $ U_{e_{\mathrm{L}}}\cdot Y_e$ has triangular form, i.e.
\begin{eqnarray}\label{eq:DiagYe}
U_{e_{\mathrm{L}}} \cdot Y_e 
=R_{23} U_{13} R_{12} P_0\cdot 
 \left(\begin{array}{ccc}
|p|\,e^{i \phi_p} & |d|\,e^{i \phi_d} & |a|\,e^{i \phi_a} \\
|q|\,e^{i \phi_q} & |e|\,e^{i \phi_e} & |b|\,e^{i \phi_b} \\
|r|\,e^{i \phi_r} & |f|\,e^{i \phi_f} & |c|\,e^{i \phi_c}
\end{array}\right)
=
\left(\begin{array}{ccc}
* & 0 & 0 \\
* & * & 0 \\
* & * & *
\end{array}\right)\!,
\end{eqnarray} 
with $P_0$ given by 
\begin{eqnarray}
P_0:= 
 \begin{pmatrix}
 1&0&0\\0&e^{i\beta^e_2}&0\\0&0&e^{i\beta^e_3}
  \end{pmatrix}
\end{eqnarray}    
and with $R_{23}(\theta_{23}),U_{13}(\theta_{13},\delta)$ and $R_{12}(\theta_{12})$ defined as in equation 
(\ref{eq:R23U13R12P0}).
From equation (\ref{eq:DiagYe}), together with 
equation (\ref{eq:SeqDominanceCL}), we obtain for the mixing angles
$\theta_{12},\theta_{13}$ and $\theta_{23}$ 
 \begin{subequations}\label{eq:mixings_general}\begin{eqnarray}
\label{eq:t12_full}\tan (\theta_{12}) &\approx& 
 \frac{
-\frac{a}{c} e^{i (\phi_a - \phi_c)} + \frac{d}{f} e^{i (\phi_d - \phi_f)}
}{
\frac{b}{c} e^{i (\phi_b - \phi_c + \beta^e_2)} - \frac{e}{f} 
e^{i (\phi_e - \phi_f + \beta^e_2)}
}\; ,\\
\label{eq:t13_full}\tan (\theta_{13}) &\approx&  -\frac{
s_{12}\,|e| \, e^{i (\phi_e + \beta^e_2+\delta)}
+c_{12}|d| \, e^{i (\phi_d + \delta)} 
}{
|f|\,e^{i (\phi_f + \beta^e_3)}}  \vphantom{\frac{f}{f}}\; , \vphantom{\frac{f}{f}}\\
\label{eq:t23_full}\tan (\theta_{23}) &\approx&   \frac{  s_{12}\, |a| \,e^{i \phi_a} -  c_{12} \,|b| \,e^{i (\phi_b+\beta^e_2)}}{
c_{13}\,|c|\,e^{i (\phi_c+\beta^e_3)} - 
s_{13}\, (c_{12} \, |a| \, e^{i (\phi_a + \delta)} + s_{12} \, |b| \, e^{i (\phi_b
+ \beta^e_2 + \delta)})
}\vphantom{\frac{f}{f}}\; ,
\end{eqnarray}\end{subequations}
where the CP phases $\beta^e_2,\beta^e_3$ and $\delta$ are determined by the
requirement that the mixing angles in equation (\ref{eq:mixings_general}) are
real and in the range $[0,\tfrac{\pi}{2}]$. The phases are thus given by
 \begin{subequations}\label{eq:phases_general}\begin{eqnarray}
\label{eq:beta2e}\beta^e_2&\approx& 
\mbox{Arg}\,\left(
\frac{
- \frac{a}{c} e^{i (\phi_a - \phi_c)} + \frac{d}{f} e^{i (\phi_d - \phi_f)}
}{
\frac{b}{c} e^{i (\phi_b - \phi_c )} - \frac{e}{f} 
e^{i (\phi_e - \phi_f )}
}
\right) ,\\
\label{eq:beta3e}\beta^e_3 - \delta &\approx& 
\mbox{Arg}\,\left(
-\frac{
s_{12}\,|e| \, e^{i (\phi_e + \beta^e_2)}
+c_{12}|d| \, e^{i \phi_d } }{
|f|\,e^{i \phi_f}}
\right) ,\\
\label{eq:delta}\delta &\approx& 
\mbox{Arg}\,\left(
\frac{  s_{12}\, |a| \,e^{i \phi_a} -  c_{12} \,|b| \,e^{i (\phi_b+\beta^e_2)}}{
c_{13}\,|c|\,e^{i (\phi_c+\beta^e_3- \delta)} - 
s_{13}\, (c_{12} \, |a| \, e^{i \phi_a } + s_{12} \, |b| \, e^{i (\phi_b
+ \beta^e_2 )})
}
\right) .
\end{eqnarray}\end{subequations}
We can extract the parameters in the following sequence: 
First we calculate $\beta^e_2$ and, using this result, we determine 
$\theta_{12}$. This allows to calculate $\beta^e_3-\delta$ and subsequently 
$\theta_{13}$. Then, $\beta^e_3- \delta$ has to be eliminated in equation (\ref{eq:delta}) in
order to obtain $\delta$. Using all the previous results allows to
calculate $\theta_{23}$. 
The mass eigenvalues of the charged leptons are approximately given by 
 \begin{subequations}\label{eq:ChargedLeptonMasses}\begin{eqnarray}
m_\tau &\approx& \left(|a|^2+|b|^2+|c|^2\right)^{\tfrac{1}{2}} v_d \;,\\ 
m_\nu &\approx& 
 \left(|d|^2+|e|^2+|f|^2 - 
 \frac{|d^* a + e^* b + f^* c|^2}{m^2_\tau}
 \right)^{\tfrac{1}{2}} v_d \;,\\ 
m_e &\approx& {\cal O} \left(|p|,|q|,|r|\right) \, v_d  \vphantom{\frac{f}{b}}\; .
 \end{eqnarray}\end{subequations}
The mixing angle $\theta_{13}$ has a present experimental upper bound of roughly 
$15^\circ$. From equation (\ref{eq:t13_full}), we see that a natural possibility 
for obtaining a small $\theta_{13}$ is\footnote{
If $a/b = d/e$, 
which implies that the two zero entries in the first row of 
equation (\ref{eq:DiagYe}) can be achieved by a
$R_{12}$-rotation alone and thus no $U_{13}$ rotation is required,   
$\theta_{13}$ is approximately zero. This condition, which is equivalent to 
$a \cdot e = d \cdot b$, has been used in the model 
proposed in \cite{Altarelli:2004jb,Romanino:2004ww}. 
Note that our condition (\ref{eq:CondForSmallT13}) does not in general 
satisfy the condition $a \cdot e = d \cdot b$.
}
\begin{eqnarray}\label{eq:CondForSmallT13}
|d|,|e| \ll |f| \; .
\end{eqnarray} 
In leading order in $|d|/|f|$ and $|e|/|f|$, the formulae of equation 
(\ref{eq:mixings_general}) and
(\ref{eq:phases_general}) simplify considerably. 
For the mixing angles $\theta_{12},\theta_{23}$ and $\theta_{13}$, we obtain   
\begin{subequations}\label{eq:mixings_1b}\begin{eqnarray}
\label{eq:t12_C2}\tan (\theta_{12}) &\approx&  \frac{|a|}{ |b|} \; ,\\
\label{eq:t23_C2}\tan (\theta_{23}) &\approx&   \frac{  s_{12}\, |a| +  c_{12} \,|b| }{
|c| } \;,\vphantom{\frac{f}{f}}\\
\label{eq:t13_C2}\tan (\theta_{13}) &\approx&  \frac{
s_{12}\,|e| \, e^{i (\phi_a - \phi_b + \phi_e +\delta)} - 
c_{12}|d| \, e^{i (\phi_d + \delta)} 
}{
|f|\,e^{i (\phi_a - \phi_c +  \phi_f)}}  \vphantom{\frac{f}{f}}\; ,
\end{eqnarray}\end{subequations}
where the Dirac CP phase $\delta$ is determined such that $\theta_{13}$ 
is real, which requires 
\begin{eqnarray}\label{eq:mixings_1c}
\label{eq:delta_C2}\tan (\delta) &\approx& 
\frac{c_{12}\, |d| \,\sin (\phi_a - \phi_c - \phi_d + \phi_f) - s_{12}\, |e|\,
\sin (\phi_b - \phi_c - \phi_e + \phi_f)}{
c_{12} \,|d| \,\cos (\phi_a - \phi_c - \phi_d + \phi_f) - s_{12}\, |e| \,\cos
(\phi_b - \phi_c - \phi_e + \phi_f)} 
\; .\vphantom{\frac{f}{f}}
\end{eqnarray}
Given $\tan (\delta)$, $\delta$ has to be chosen such that 
$\tan(\theta_{13})\ge 0$ in order to match with the usual convention
$\theta_{13}\ge 0$. 
The phases $\beta^e_2$ and $\beta^e_3$ from the charge lepton
sector are given by 
 \begin{subequations}\label{eq:mixings_1a}\begin{eqnarray}
\label{eq:b2_C2}\beta^e_2 &\approx& \phi_a - \phi_b + \pi \; , \\
\label{eq:b3_C2}\beta^e_3 &\approx& \phi_a - \phi_c \; . 
\end{eqnarray}\end{subequations}
Note that in the case that the neutrino sector induces Majorana phases, the 
total Majorana phases $\beta_2$ and $\beta_3$ of the MNS matrix are 
given by 
 \begin{subequations}\label{eq:mixings_1d}\begin{eqnarray}
\label{eq:total_b2_C2}\beta_2 &\approx& \beta^e_2 + \beta^\nu_2 \; ,\\
\label{eq:total_b3_C2}\beta_3 &\approx& \beta^e_3 + \beta^\nu_3 \; .
\end{eqnarray}\end{subequations} 
$\theta_{13}$ only depends on $d/f$ and $e/f$ from 
the Yukawa couplings to the subdominant right-handed muon and on $\theta_{12}$.  
We find that in the limit $|d|,|e| \ll |f|$, the two large mixing angles 
$\theta_{12}$ and $\theta_{23}$ approximately depend only on 
$a/c$ and $b/c$ from 
the right-handed tau Yukawa couplings.  
Both mixing angles are large if $a,b$ and $c$ are of the same order.

\section{A Model with SO(3) Flavour Symmetry and Real Vacuum Alignment}\label{sec:C2}
 
Spontaneously broken SO(3) flavour symmetry 
and a real alignment mechanism for the 
SO(3)-breaking vevs \cite{Antusch:2004xd}
naturally allows for 
sequential dominance for the right-handed charged leptons as well as for 
the right-handed neutrinos.  
It furthermore leads to three zero entries in the neutrino and charged 
lepton Yukawa matrices in a special basis. 
The motivation for choosing such a basis would come from a full theory 
 beyond the scope of this framework.
D-brane models leading to textures with a dominant third column 
have been discussed e.g.~in 
\cite{Everett:2000up}. 
 A classification of the textures arising from SO(3)-breaking with 
 real vacuum alignment 
 can be found in table 4 of \cite{Antusch:2004xd} and a particular interesting
 class of models there is type C2, in which 
 the neutrino and charged lepton Yukawa matrices have the form   
\begin{eqnarray}
Y_e =\left(\begin{array}{ccc}
0&0&a\,e^{i \delta_3}   \\
q\,e^{i \delta_1}&0&b\,e^{i \delta_3} \\
r\,e^{i \delta_1}&f\,e^{i \delta_2}&c\,e^{i \delta_3}
\end{array}
\right), \;\;\;
Y_\nu =\left(\begin{array}{ccc}
0&0&a'\,e^{i \delta'_3}   \\
q'\,e^{i \delta'_1}&0&b'\,e^{i \delta'_3} \\
r'\,e^{i \delta'_1}&f'\,e^{i \delta'_2}&c'\,e^{i \delta'_3}
\end{array}
\right), 
\end{eqnarray}
with real parameters $\{a,b,c,f,q,r,\da,\db,\dc\}$ and
$\{a',b',c',f',q',r',\daP,\dbP,\dcP\}$.  

\subsection{The Charged Lepton Sector}\label{sec:C2_lept}
In the charged lepton sector the sequential dominance condition  
\begin{eqnarray}
|a|,|b|,|c| \gg |f| \gg  |q|,|r|  
\end{eqnarray}
which is similar to equation (\ref{eq:SeqDominanceCL}) leads to 
the desired hierarchy among the charged lepton masses and to 
small right-handed mixing angles of $U_{e_{\mathrm{R}}}$.
A characteristic feature of the Yukawa matrices, which stems from real 
alignment of the SO(3)-breaking vacuum, is that 
 each column of the Yukawa matrices has a common complex phase 
 $\delta_I$ or $\delta'_I$ ($I\in\{1,2,3\}$). 
  The lepton mixing angles from the charged lepton 
  sector can be extracted using equation 
(\ref{eq:mixings_1b}), which reduces to 
\begin{subequations}\label{eq:mixings_2}\begin{eqnarray}
\label{eq:t12_2}\tan (\theta^e_{12}) &\approx&  \frac{|a|}{|b|} \; ,\\
\label{eq:t13_2}\tan (\theta^e_{13}) &\approx& 0 \vphantom{\frac{a}{a}}\; ,\\
\label{eq:t23_2}\tan (\theta^e_{23}) &\approx&  \frac{\sqrt{|a|^2+|b|^2}}{|c|} .  
\end{eqnarray}\end{subequations}  
Approximately zero mixing $\theta^e_{13}$ is a consequence of the two zero  
Yukawa couplings $(Y_e)_{12}$ and $(Y_e)_{22}$ 
to the right-handed muon. A texture for $Y_e$ with a 
structure similar to the one we have found from spontaneous SO(3)-breaking 
has been considered in \cite{Babu:2001cv}. 
Note that due to the specific structure of $Y_e$, the complex phases 
$\delta'_I$ can be absorbed by $U_{e_\mathrm{R}}$ and they thus have no effect
on the CP phases of the MNS matrix. With zero $\theta_{13}$ at leading order, 
the Dirac CP phase $\delta$ is undefined. We now turn to the parameters of the
neutrino sector and to corrections from small mixing in the neutrino sector for
the lepton mixing angles and for the Dirac CP phase. 
 
\subsection{The Neutrino Sector}\label{sec:C2_nu}  
In leading order, the Majorana mass matrix of the heavy right-handed neutrinos 
has a diagonal structure from SO(3)-symmetry,  
\begin{eqnarray}
M_\mathrm{RR}
= 
\left(\begin{array}{ccc}
\!X&0&0\!\\
\!0&Y&0\!\\
\!0&0&X'\!
\end{array}
\right)\!.
\end{eqnarray}
In order to obtain small lepton 
mixings from the neutrino sector, we impose the sequential RHND condition  
\begin{eqnarray}\label{eq:SequSubDominace}
\frac{|f'|^2}{Y}\gg 
\frac{|q'|^2,|r'|^2}{X} \gg
\frac{|a'|^2,|b'|^2,|c'|^2}{X'}
\; .
\end{eqnarray}
The dominant and the subdominant columns of $Y_\nu$ are equivalent 
to the form of equation (\ref{eq:SmallMixingFromMnu}). 
The type I version of the model implies a strongly hierarchical neutrino mass spectrum.
  We can however generalize the above type I see-saw model to a 
  type II see-saw 
 model (see e.g.~\cite{Lazarides:1981nt,Mohapatra:1981yp}) where the neutrino
 mass matrix is given by 
\begin{eqnarray}
m^\nu_{\mathrm{LL}} = m^\mathrm{II}_{\mathrm{LL}} +
m^\mathrm{I}_{\mathrm{LL}}\;.
\end{eqnarray}
$m^\mathrm{II}_{\mathrm{LL}}$ corresponds to an 
additional direct mass term for the neutrinos. The SO(3) flavour symmetry 
forces the direct mass term to be proportional to 
the unit matrix
\begin{eqnarray}
m^{\mathrm{II}}_{\mathrm{LL}} \approx m^{\mathrm{II}}\, 
\left(\begin{array}{ccc}
1&0&0\\
0&1&0\\
0&0&1
\end{array}\right)
\end{eqnarray}
 at leading order in the effective theory. 
This allows to upgrade the type I scenario of section
\ref{sec:C2} to natural models for hierarchical well as for partially 
degenerate neutrino masses \cite{Antusch:2004xd}. The neutrino mass matrix is
approximately given by
\begin{eqnarray}
m^{\nu}_{\mathrm{LL}} =
\left(\begin{array}{ccc}
m^{\mathrm{II}}&0&0\\
0&m^{\mathrm{II}} - |q'|^2\,\frac{ v_u^2}{X} \,e^{i 2\daP}& -|q'|\,|r'|\,\frac{ v_u^2}{X}\,e^{i 2\daP}\\
0&-|q'|\,|r'|\,\frac{ v_u^2}{X}\,e^{i 2\daP}&m^{\mathrm{II}}- |f'|^2\,\frac{
v_u^2}{Y} \,e^{i 2\dbP}- |r'|^2\,\frac{ v_u^2}{X} \,e^{i 2\daP}
\end{array}\right),
\end{eqnarray}
where we have only considered $m^\mathrm{II}_{\mathrm{LL}}$ and the dominant and the subdominant part of
$m^\mathrm{I}_{\mathrm{LL}}$. 
From the neutrino sector, we obtain corrections to the total lepton
mixings due to a small mixing $\theta^\nu_{23}$ of $U^\dagger_{\nu_\mathrm{L}}$
determined by 
\begin{eqnarray}\label{eq:mixings_nu}
\label{eq:t23_nu}\theta^\nu_{23} &\approx& 
 \frac{\mbox{sign}(q' r')\,|q'| \, |r'|\,\frac{ v_u^2}{X} 
}{
2 m^{\mathrm{II}}\,\sin(2 \daP)\,\sin (\delta)
+ |f'|^2\,\frac{v_u^2}{Y} \,\cos(2\daP-2\dbP+ \delta)}\; , \vphantom{\frac{|a|}{|b|}}
\end{eqnarray}
where $\delta$ is given by
\begin{eqnarray}\label{eq:deltatilde_nu}
\label{eq:t23_nu} \tan(\delta) &\approx& 
 \frac{
|f'|^2\,\frac{v_u^2}{Y}\,\sin(2\daP-2\dbP)
}{2 m^{\mathrm{II}}\,\cos(2 \daP ) -
 |f'|^2\,\frac{v_u^2}{Y}\,\cos(2\daP-2\dbP)
}
 \vphantom{\frac{|a|}{|b|}}
\end{eqnarray}
and, given $\tan (\delta)$, $\delta$ is chosen such that $\theta^\nu_{23}>0$.  
The latter leads to a positive 
induced total lepton mixing angle $\theta_{13}$. 
$\delta$ is the Dirac CP phase, which is induced from the complex
phases in the neutrino
sector. 
The smallness of  $\theta^\nu_{23}$ is guaranteed by equation 
(\ref{eq:SequSubDominace}). 
Including the corrections from the neutrino sector in leading order in $\theta^\nu_{23}$, 
the total mixing angles of the MNS matrix are given by  
 \begin{subequations}\label{eq:mixings_total}\begin{eqnarray}
\label{eq:t12_c}\tan(\theta_{12}) &\approx&
\vphantom{\frac{a}{a}}\tan(\theta^e_{12})
 \; ,\\
\label{eq:t13_c}\theta_{13} &\approx& \theta^\nu_{23}\,
\vphantom{\frac{a}{a}} \sin (\theta^e_{12}) 
\; ,\\
\label{eq:t23_c}\tan(\theta_{23}) &\approx& 
\frac{\sin(\theta^e_{23})+\theta^\nu_{23}\, \cos (\theta^e_{12}) \,
\cos(\theta^e_{23}) \cos(\delta) }{
\cos(\theta^e_{23})-\theta^\nu_{23}\,\cos (\theta^e_{12}) \,
\sin(\theta^e_{23})  \cos(\delta)
}\; .  
\end{eqnarray}\end{subequations}  
 The neutrino mass eigenvalues of the type I see-saw version of the scenario 
are given by 
\begin{subequations}\begin{eqnarray}
\label{eq:m1I_C2}m_1^{\mathrm{I}} &=&  
{\cal O} \left(\frac{|a'|^2, |b'|^2, |a'| |b'|}{X'}\,v_u^2\right)   \;\approx\;0 
\; , \vphantom{\frac{|e|}{|f|}}\\
\label{eq:m2I_C2} m_2^{\mathrm{I}} &\approx& \frac{|q'|^2 }{X}\,v_u^2\;,\\
\label{eq:m3I_C2}m_3^{\mathrm{I}} &\approx& 
\frac{|f'|^2}{Y}\,v_u^2\;. 
\end{eqnarray}\end{subequations} 
The mass eigenvalues of the complete type II neutrino mass matrix are given by
 \begin{subequations}\label{eq:ComplMassEigenvOfMnuTypeII_C2}\begin{eqnarray}
m_1  &\approx&  |m^{\mathrm{II}}|\;,\\
m_2  &\approx& | m^{\mathrm{II}} - m_2^{\mathrm{I}} \,e^{i 2 \da}|\;,\\
m_3  &\approx& | m^{\mathrm{II}} - m_3^{\mathrm{I}} \,e^{i 2 \db}|\;. 
\end{eqnarray} \end{subequations} 
For $m^{\mathrm{II}} \not= 0$, the Majorana phases $\beta_2$ and $\beta_3$ 
can be extracted by
\begin{subequations}\label{eq:MajPhases_C2}\begin{eqnarray}
\beta_2  &\approx& \frac{1}{2}\mbox{arg}\,( m^{\mathrm{II}} - m_2^{\mathrm{I}} \,e^{i 2 \da})\;,\\
\beta_3  &\approx& \frac{1}{2}\mbox{arg}\,( m^{\mathrm{II}} - m_3^{\mathrm{I}} \,e^{i 2 \db})\;.
\end{eqnarray}\end{subequations} 
 The corrections from the leading order neutrino mass matrix in
general result in a non-zero value for $\delta$ given by equation 
(\ref{eq:deltatilde_nu}). The additional corrections for    
$\theta_{13}$, which we will discuss below, could also 
modify the value for the Dirac CP phase.

\section{The Neutrino Mass Scale and Predictions for
$\boldsymbol{\theta_{13}}$}\label{sec:theta13}
A prediction $\theta_{13} \approx 0$ from the charged lepton sector  
as in section \ref{sec:C2_lept} will 
be subject to corrections from e.g.~next-to-leading order effective 
operators allowed by symmetry, from the small mixings of 
$U_{\nu_{\mathrm{L}}}$,  
and from the renormalization group 
(RG) running between the high energy scale where the
models are defined and the low energy scale where experiments 
are performed \cite{Antusch:2003kp}. 
The corrections to $\theta_{13} \approx 0$ are linked to the 
neutrino mass scale. 
We expect the corrections from 
next-to-leading order effective operators to be larger for a larger neutrino 
mass scale since the parameters of
the neutrino mass matrix are more sensitive to small modifications for a larger
mass scale. 
Because of the latter reason, the RG effects are generically 
enhanced for a larger neutrino mass scale as well.  
However, for $\theta_{13}$, this is compensated in the type-II-upgrade 
scenarios \cite{Antusch:2004xd} by the fact that the Majorana phase $\beta_2$ 
gets smaller with increasing neutrino mass scale, 
which has a damping effect on the running 
of $\theta_{13}$ \cite{Antusch:2003kp}. 
Using sequential right-handed neutrino dominance for the type I part 
 $m^{\mathrm{I}}_{\mathrm{LL}}$ of the neutrino mass matrix, a small mixing 
 $\theta^{\nu}_{23}= {\cal O} (m_2^{\mathrm{I}}/m_3^{\mathrm{I}})$ 
 is generated from the neutrino mass matrix, as pointed out in section 
 \ref{sec:SmallNuMixing}. 
 This induces a contribution $\Delta \theta_{13}$ to the total lepton mixing 
 $\theta_{13}$ given by $\Delta \theta_{13} \approx \theta^{\nu}_{23} \,
 \sin(\theta_{12})$.     
For type I see-saw scenarios, this results in a correction 
$\Delta \theta_{13} \lesssim  5^\circ$. 
Since in the type-II-upgrade scenarios \cite{Antusch:2004xd} 
$m_2^{\mathrm{I}}/m_3^{\mathrm{I}}$ decreases with increasing 
$m^{\mathrm{II}}$, this correction will get smaller for a larger neutrino 
mass scale. For the model discussed in section \ref{sec:C2}, $\theta^{\nu}_{23}$
is given by equation (\ref{eq:t13_c}).

\section{Conclusions}
We have investigated the possibility that bi-large lepton mixing 
originates from the charged lepton sector 
in models with sequential dominance for the right-handed 
charged leptons as well as for right-handed neutrinos. 
In the charged lepton sector, sequential dominance implies the desired hierarchy for the charged lepton masses and in the 
neutrino sector, it can naturally lead to 
small mixings from the neutrino mass matrix even if the neutrino Yukawa matrix
has a similar structure to the charged lepton Yukawa matrix.
We have derived analytical expressions for the mixing 
angles and CP phases of the lepton mixing matrix $U_{\mathrm{MNS}}$ 
for this case. 
We find that for small $\theta_{13}$, the two large mixing angles $\theta_{12}$ and 
$\theta_{23}$ are determined by the Yukawa couplings to the 
dominant right-handed tau lepton. The mixing angle $\theta_{13}$ is then governed
 by the subdominant right-handed muon Yukawa couplings. 
We find a condition for naturally small $\theta_{13}$, 
which is different from the vanishing determinant condition of 
\cite{Altarelli:2004jb,Romanino:2004ww}.
We have discussed an example with sequential dominance 
realized within the framework of spontaneously broken SO(3) flavour symmetry.  
Real vacuum alignment can lead to additional zero 
entries in the lepton Yukawa matrices which predict small 
$\theta_{13}$.  
Type I see-saw models of this type can be upgraded to type II see-saw models 
where hierarchical neutrino masses are as natural a partially degenerate 
ones \cite{Antusch:2004xd}.   
We have discussed the predictions for $\theta_{13}$ and corrections to it, 
depending on the neutrino mass scale. Although the mixing $\theta_{13}$
generated in the charged lepton sector is very small, corrections to it from 
next-to-leading order effective operators, small mixings from the neutrino mass
matrix and renormalization group running could lift the prediction to 
$\theta_{13} \lesssim 5^\circ$ 
 or $\sin^2 (2 \theta_{13})\lesssim 3\cdot 10^{-2}$,   
 which might lie within reach of planned reactor experiments 
(see e.g.~\cite{reactor}). 
 In the
type-II-upgrade scenarios, $\theta_{13}$ is in general 
predicted to be smaller for a larger neutrino mass scale.

\section*{Acknowledgements}
We acknowledge support from the PPARC grant PPA/G/O/2002/00468.

\providecommand{\bysame}{\leavevmode\hbox to3em{\hrulefill}\thinspace}

\end{document}